\newcommand\pcc{{\rm cm^{-3}}}

\newcommand\kelvin{{\rm K}}
\newcommand\kmps{{\rm km\,s^{-1}}}
\newcommand\erg{{\rm erg}}
\newcommand\yr{{\rm yr}}

\newcommand\Msol{{M_\odot}}

\newcommand\XHCN{[{\rm HCN}]/[{\rm CO}]}

\newcommand\Tkin{{T_{\rm kin}}}
\newcommand\nH{{n_{\rm H}}}
\newcommand\ncrit{{n_{\rm crit}}}
\newcommand\Column[1]{{N_{#1}}}
\newcommand\dColumn[1]{{\Column{#1}/{{\rm d}v}}}
\newcommand\vlsr{{v_{\rm LSR}}}
\newcommand\Tmb{{T_{\rm MB}}}
\newcommand\Ta {{T_{\rm A}^*}}

\newcommand\CO[2]{{\rm {^{#1}C{^{#2}O}}}}
\newcommand\HCN[1]{{\rm H{^{#1}C}N}}
\newcommand\HCOp[1]{{\rm H{^{#1}C}O^+}}
\newcommand\SiO[2]{{\rm {^{#1}Si{^{#2}O}}}}

\newcommand\JJ[2]{\mbox{{\it J}=#1\mbox{--}#2}}
\newcommand\thecomplex{{\it l}=1.3^\circ}

\documentstyle[twocolumn]{pasj00}
\draft

\SetRunningHead{Tanaka et al.}{High Resolution Mappings of The $L=1.3^\circ$ Complex in Molecular Lines}
\title{HIGH RESOLUTION MAPPINGS OF THE $L=1.3^\circ$ COMPLEX IN MOLECULAR LINES : DISCOVERY OF A PROTO-SUPERBUBBLE}
\author{Kunihiko Tanaka$^1$, Kazuhisa Kamegai$^1$, Makoto Nagai$^2$, and Tomoharu Oka$^2$}
\affil{%
   $^1$Institute of Astronomy, Graduate School of Science, University of Tokyo, 2--21--1 Osawa,\\
   Mitaka-shi, Tokyo 181--8588}
\email{tanaka@ioa.s.u-tokyo.ac.jp}
\affil{%
   $^2$Department of Physics, Graduate School of Science, University of Tokyo, 7--3--1 Hongo,\\
   Bunkyo-ku, Tokyo 113--0033}

\begin{document}
\maketitle

\begin{abstract}
We report the results of molecular line observations toward the $l=1.3^\circ$ complex, an anomalous cloud complex in the central molecular zone of the Galaxy.   
We have taken high resolution maps of the CO $\JJ{1}{0}$, HCN $\JJ{1}{0}$, $\HCOp{}$, SiO $\JJ{1}{0}$ and $\JJ{2}{1}$ lines.  
The complex is found to be rich in shells and arcs of dense molecular gas.  
We have identified 9 expanding shells in HCN maps and compact SiO features associated to the shells.  
The intensity ratios of HCN/CO, $\HCOp{}{}$/CO and CO $\JJ{3}{2}$/$\JJ{1}{0}$ are coherently enhanced by a factor of a few in gas with an LSR velocity higher than $110\ \kmps$.  
The high-velocity gas has a high density ($\nH\sim10^{4.5}\ \pcc$) and high SiO/$\CO{13}{}$ intensity ratio indicating that the gas was shocked.  
The typical HCN/$\HCOp{}$ intensity ratio is found to be 2.3, being higher by an factor of a few than those in the Galactic disk clouds.  
The typical kinetic energy and expansion time of the shells are estimated to be $10^{50.9\mbox{--}52.5}$ erg and $10^{4.6\mbox{--}5.3}$ yr, respectively.  
The kinetic energy could be furnished by multiple supernova and/or hypernova explosions with a rate of $10^{-3\mbox{--}-4}$ yr$^{-1}$.  
These estimates suggest that the expanding shells as a whole may be in the early stage of superbubble formation.  
This proto-superbubble may be originated by a massive cluster formation which took place $10^{6.8\mbox{--}7.6}\ \yr$ ago.
\end{abstract}

\section {INTRODUCTION}

\begin{table*}[t] \begin{center}
\caption{OBSERVED LINES}\label{Tab1}
\pagestyle{empty}
\small
\begin{tabular}{llcccccc}
\hline\hline
\multicolumn{2}{l}{transition} & 
frequency & 
\hspace{-2mm}beamsize\hspace{-2mm} &
grid &
mapped\ area &
{1/$\eta_{\rm MB}$} &
\\
 & & 
{\footnotesize(GHz)} & 
{\footnotesize($''$)} &
{\footnotesize($''$)} &
{\footnotesize($'\times'$)} &
\\
\hline
CO                     & $\JJ{1}{0}$ & 115.27120 & 15 & 20.55 & $14.4 \times 28.4$ & 2.2 \\
HCN      \hspace{-3mm} & $\JJ{1}{0}$ &  88.63    & 19 & 20.55 & $14.4 \times 28.4$ & 2.5 \\
$\HCOp{}$\hspace{-3mm} & $\JJ{1}{0}$ &  88.69751 & 19 & 20.55 & $14.4 \times 28.4$ & 2.5 \\
SiO                    & $\JJ{1}{0}$ &  43.42376 & 38 & 82.2  & $13.7 \times 13.7$ & 1.3 \\
                       & $\JJ{2}{1}$ &  86.84696 & 18 & 82.2  & $13.7 \times 13.7$ & 1.5 \\
\hline
\end{tabular}

\end{center} 
\end{table*}

It has long been recognized that the central molecular zone (CMZ) of our Galaxy is characterized by high gas kinetic temperature ($\sim 30-60$ K; \cite{mor83}) above the dust temperature ($\sim$ 30 K; \cite{cl89,pp00}), and by high gas density ($\sim10^4\ \pcc$) as well.  
Molecular clouds there show evidence of a highly turbulent nature, with large velocity widths ($\sim 50\ \kmps$).  
It is argued that dissipation of supersonic turbulence is the dominant gas heating mechanism in the CMZ.  
The presence of the pervasive shocks is suggested by widespread thermal SiO emission \citep{mpi96, hue98}.  
High resolution surveys of molecular lines and dust continuum at millimeter/submillimeter wavelengths \citep{oka98, tsu98, pp00} show that the CMZ contains a number of filaments, arcs and shells, which are indicative of local turbulent sources.

Recently we made a survey of the CMZ in the submillimeter CO $\JJ{3}{2}$ line (345.7955 GHz) covering $-1.5^\circ<l<+1.0^\circ$, $-0.2^\circ<b<+0.2^\circ$ by using the ASTE 10\ m Telescope (\cite{pap1}, hereafter Paper I).  
The CO $\JJ{3}{2}$ rotational transition has a relatively large critical density ($\ncrit \sim 10^4\ \pcc$) and high upper sate energy ($\sim 34\ \kelvin$), which allow us to trace shocked molecular gas easily.  
The most important discovery of the survey is that a number of clumps with enhanced $\JJ{3}{2}$/$\JJ{1}{0}$ ratio ($>1.5$) are widely distributed in the CMZ.  
These clumps are suspected as spots where interaction with local turbulent sources is taking place, although the turbulent sources are unidentified in many cases.  

One of the prominent high CO $\JJ{3}{2}$/$\JJ{1}{0}$ ratio clumps is CO 1.27+0.01 in the $\thecomplex$ complex (also referred as '$l=1.5^\circ$ complex'; \cite{bar88, oka01}), an unusual molecular feature with broad-velocity-width and a large latitudinal scale height.  
Especially the high ratio gas highlights two expanding shells (the `major' and `minor' shells; \cite{oka01}) included in the complex.  
The $\thecomplex$ complex also includes a cloud with enhanced SiO abundance ($[{\rm SiO}]/[{\rm H}]\sim10^{-8}$; \cite{hue98}).  
The origin of the disturbed morphology and kinematics in the $\thecomplex$ complex has been discussed in the context of the star formation history and large scale gas kinematics in the CMZ.  
Each of the expanding shells in CO 1.27+0.01 is considered to be formed by multiple supernova explosions or by a hypernova explosion \citep{oka01}.  
Hu\"ettemeister et al. (1998) argues that the high SiO abundance in the complex is due to local explosive events and/or cloud collision related to the large scale gas motion in the bar potential.

In this paper, we report the results of follow-up observations to the ASTE CO $\JJ{3}{2}$ survey, high resolution mappings of the $\thecomplex$ complex in the CO, $\HCN{}$, $\HCOp{}$ and SiO rotational lines.  
The rotational transitions of $\HCN{}$ and $\HCOp{}$ molecules are sensitive to higher density ($n_{\rm crit}\sim 10^5\ \pcc$) than the CO $\JJ{3}{2}$ line.   
The $\HCN{}$ and $\HCOp{}$ mappings have been carried out with higher angular resolution, $\sim 20''$, than the ASTE survey.  
The SiO molecule is an established tracer of shock-heated molecular clouds, as it is formed via chemical reactions initiated by evacuation of depleted Si into the gas-phase \citep{dow82,zfi89}.  
We investigate detailed morphology and kinematics of the high density gas in the complex with these data.

In the following section we describe the observations.  
The acquired data sets are presented in Section 3, and morphology and kinematics of the complex are analyzed in detail.  
In Section 4, we estimate physical conditions and kinetic energy of turbulent gas, based on which we argue that the $\thecomplex$ complex contains a kind of molecular superbubble.  

\section{OBSERVATIONS}
The observations were made in January 2006, by using the Nobeyama Radio Observatory 45\ m telescope.
We used the BEARS (25 Beam Receiver Array System) for observations of the CO $\JJ{1}{0}$, HCN $\JJ{1}{0}$ and $\HCOp\,\JJ{1}{0}$ lines.  
The SiO $\JJ{1}{0}$ and $\JJ{2}{1}$ data were taken simultaneously by using the S40 and S100 receivers.  
Observed frequencies and respective angular resolutions are listed in Table \ref{Tab1}.  
The digital backend was operated in wide-band mode, with 0.5 MHz channel widths.  
The acquired spectra were then smoothed by the $2\ \kmps$ FWHM Gaussian and regridded onto each $2\ \kmps$ bin for convenience in comparison between the different lines.  
Antenna temperatures were calibrated by the chopper-wheel method.  
The $\Tmb$/$\Ta$ scaling factor (1/$\eta_{\rm MB}$) adopted for each line is listed in Table\ref{Tab1}.

We mapped a $14.4'\times 28.4'$ area roughly centered at CO1.27+0.01 in CO, HCN and $\HCOp{}$ lines.  
The grid spacing was $20''.55$.  
The mapped area of the SiO lines was smaller, $13'.7\times 13'.7$ and the grid spacing was $82''.2$.  
Antenna pointing accuracy was maintained within $3''$ by observing SiO $J=1-0, v=1,2$ maser lines toward VX-Sgr.   
All the observations were carried out by position switching mode.  The reference position was taken at ($l$, $b$) = ($1^\circ, -1^\circ$).

\section{RESULTS}
\subsection{Overall Morphology : Expanding Shells}
Figure \ref{Fig1} shows velocity channel maps of the CO, HCN and $\HCOp{}{}$\ $\JJ{1}{0}$ lines at $20\ \kmps$ intervals.
The CO $\JJ{3}{2}$ data taken from \citet{pap1} are also displayed in the figure, though the angular resolution is lower ($34''$ grid) and the area coverage is limited.

The maps exhibit highly complicated structure having a number of emission cavities. %
Three large shells are denoted as A, B and C in the figure, which are approximately aligned vertically to the Galactic plane.
The shell A is considered as the same structure as the ``major'' shell, which is one of the two expanding shells found in the CO $\JJ{1}{0}$ survey \citep{oka98}.  
The position of the shell C is close to another known shell, the ``minor shell'', but it offsets by $\sim 2'$ in the Galactic east.  
The shell B was unidentified in the previous observations.  In addition to the three large shells, we see a number of smaller shells in the maps.  
Identification of the shells and close look into each shell will be described in the next subsection.  

The shells/arcs are clearly seen especially in the high velocity range ($\vlsr > 110 \kmps$).  
The spatial distributions of the CO, HCN and $\HCOp{}$ emissions are similar in the high velocity range, and the shells A and C are easily recognizable in all three lines.  
The CO distribution becomes featureless in the low velocity range ($\vlsr < 110\ \kmps$), while the HCN and $\HCOp{}$ distributions are still highly fragmented.  
The shell B, which are seen in the HCN and $\HCOp{}$ maps, is not recognizable in the CO map.

The difference between the high and low velocity gas appears also in line intensity ratios.  
Figure \ref{Fig2} displays the intensity ratios of CO $\JJ{3}{2}$, HCN $\JJ{1}{0}$ and $\HCOp{}\,\JJ{1}{0}$ to the CO $\JJ{1}{0}$ and SiO\ $\JJ{1}{0}$/$\CO{13}{}\,\JJ{1}{0}$ ratio on the {\it l-V} maps.  
Contour maps of the $\CO{13}{}\,\JJ{1}{0}$ line taken from \citet{oka98} are overlaid in the figure.  
The $\CO{13}{}\,\JJ{1}{0}$ contour maps show that the bulk of the gas in the complex forms an emission ridge at velocities $70 - 110\ \kmps$.  
Typical HCN/CO and $\HCOp{}$/CO ratios are 0.1 and 0.04, respectively, in the $\CO{13}{}$ emission ridge.  
On the other hand, the HCN and $\HCOp{}{}$ emission extends to $\sim 200\ \kmps$, where the HCN/CO and $\HCOp{}{}$/CO ratios become higher by a factor of $\sim 2$.  
The high ratio appears also in the lower velocity side in the region $1.25^\circ < l < 1.35^\circ$ and $-0.1^\circ < b < 0.2^\circ$.  The CO $\JJ{3}{2}$ to $\JJ{1}{0}$ and SiO $\JJ{1}{0}$/$\CO{13}{}$ ratios show a similar trend that high-velocity gas generally has high ratio.


\begin{figure*}[p]
\begin{center}
     \FigureFile(142mm,142mm){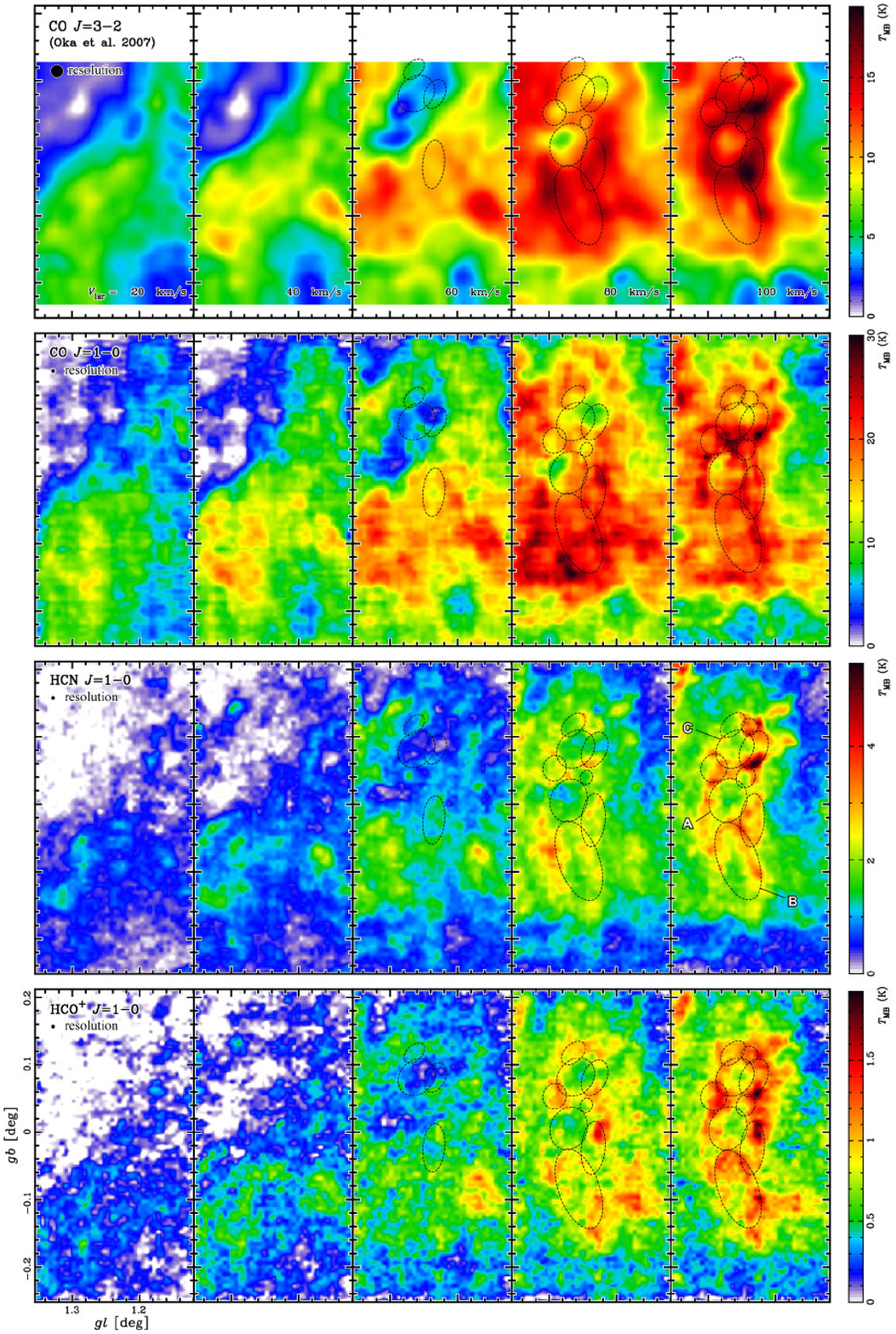}
\end{center}
\caption{Velocity channel maps of CO $\JJ{3}{2}$ \citep{pap1}, CO $\JJ{1}{0}$, HCN$\JJ{1}{0}$ and $\HCOp{}\,\JJ{1}{0}$ lines.}
\label{Fig1}
\end{figure*}

\addtocounter{figure}{-1}

\begin{figure*}[p]
\begin{center}
     \FigureFile(142mm,142mm){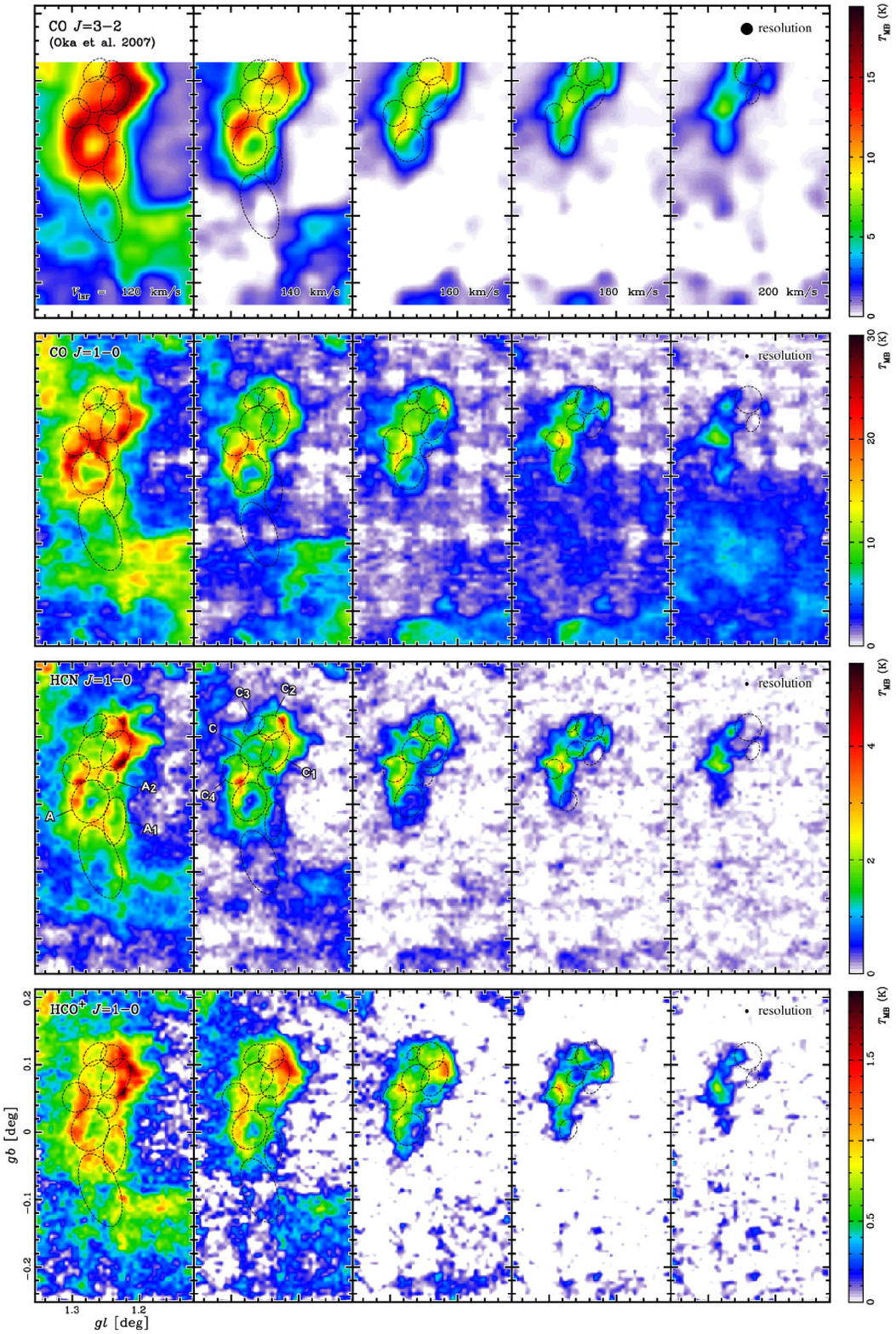}
\end{center}
 \caption{\it Contd.}
\end{figure*}


\begin{figure*}[btp]
  \begin{center}
    \FigureFile(170mm,90mm){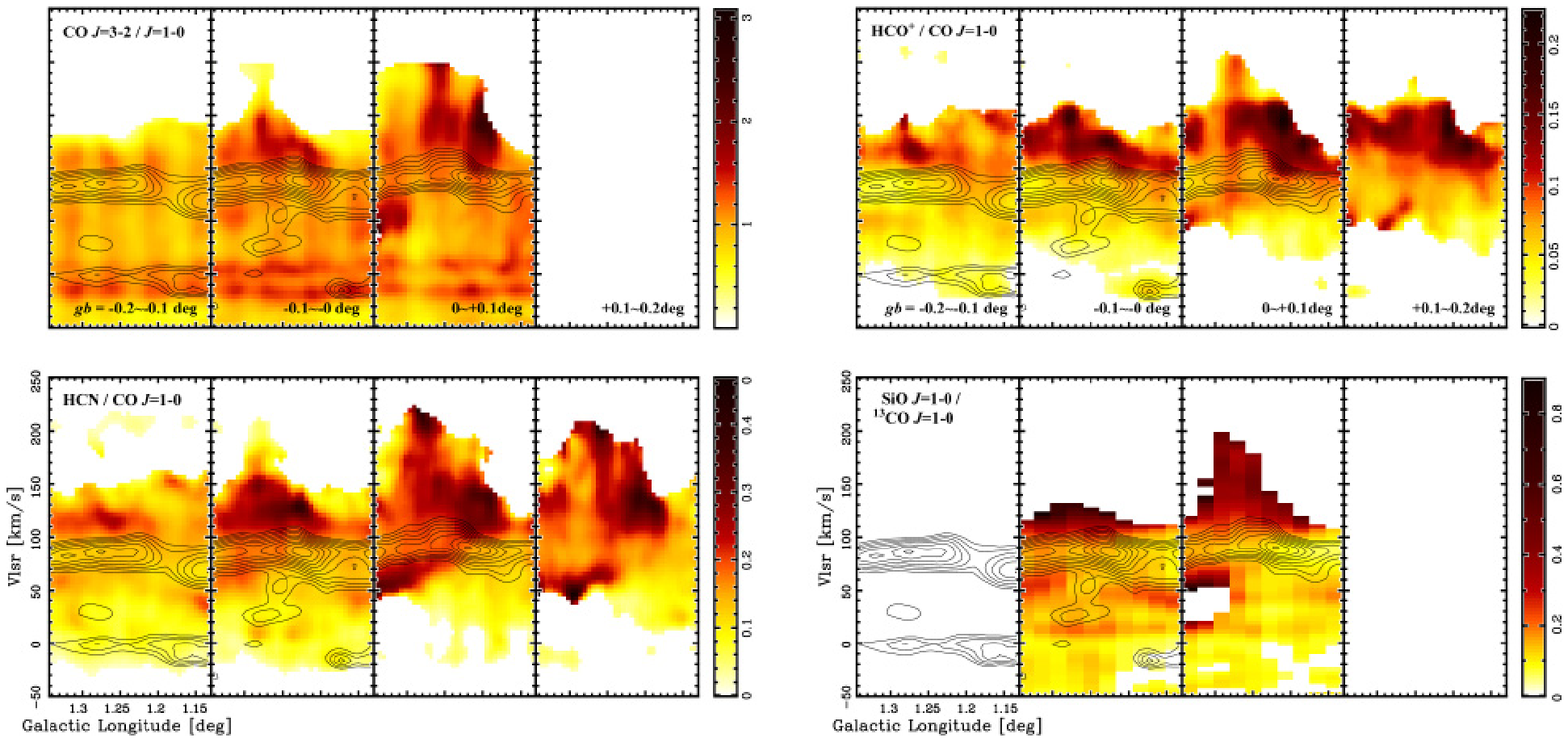}
\end{center}
\caption{
The {\it l-V} maps of ratios of the CO $\JJ{3}{2}$ (top left), HCN $\JJ{1}{0}$ (bottom left) and $\HCOp{}\,\JJ{1}{0}$ (top right) line intensities to the $\CO{}{}\,\JJ{1}{0}$ line intensity, and SiO $\JJ{1}{0}$/$\CO{13}{}\,\JJ{1}{0}$ intensity ratio (bottom right) .  
The ratios are calculated for pixels where the both line intensities are above $3\sigma$.  
In order to improve the S/N ratio, all intensity data were convolved with $41''.1$ Gaussian beam.  
Overlaid contour shows $\CO{13}{}\,\JJ{1}{0}$ line intensity (\cite{oka98}) smoothed to the same resolution. 
}\label{Fig2}
\end{figure*}


\subsection{Expanding Shells and Isolated SiO Features}

The high resolution maps of CO, HCN and $\HCOp{}$ lines are filled with a number of shells and arcs.  
We identified 9 shells with broad-velocity-width features in the HCN and $\HCOp{}$ maps and traced their locus in the {\it l-b-V} data cubes.  
In order to estimate the size and expansion velocity, we assumed ellipsoid geometry for each shell,
\begin{eqnarray}
\label{EQN_SHELLMOTION}
\left(\frac{x - x_0}{R_x}\right)^2 +
\left(\frac{y - y_0}{R_y}\right)^2 +
\left(\frac{z - z_0}{R_z}\right)^2
&=&
1\ ,
\end{eqnarray}
where the coordinates are set so that the $z$ is parallel to the line of sight, and $x$ and $y$ are to the major and minor axes of the ellipsoid projected on the plane of the sky.
We adopted a simple assumption that the ellipsoid was proportionally expanding.  
Then the term $(z-z_0)/R_{\rm z}$ in Eq(\ref{EQN_SHELLMOTION}) can be rewritten with line-of-sight expansion velocity $v_{\rm exp}$, as $(v-v_0)/v_{\rm exp}$.  

We have listed the identified expanding shells in Table \ref{Tab2}, along with their linear sizes and expansion velocities.  
We adopted 8.5 kpc as the distance to the Galactic Center in converting angular sizes into linear sizes.  
In many cases emission at the high/low velocity end is too faint to be detected, and thus the estimated expansion velocities should include large uncertainties.  
We also listed the `minimum expansion velocity' defined as the half of the velocity range within which the emission was detected.  
Schematic drawings of the shells are displayed in Fig. \ref{Fig1}.  


\begin{table}[bt]
\caption{EXPANDING SHELLS}\label{Tab2}
\begin{center}
\pagestyle{empty}
\small
\begin{tabular}{lccccccc}
\hline\hline
 & 
$l$ & 
$b$ &
$R_{\rm x} \times R_{\rm y}$ &
PA &
$\vlsr$ &  
$v_{\rm ex}$ \\
 &  
{\footnotesize(deg}) & 
{\footnotesize(deg}) &
{\footnotesize(pc}) &
{\footnotesize(deg}) &
{\hspace{-2mm}\footnotesize($\kmps$})\hspace{-2mm} &  
{\hspace{-2mm}\footnotesize($\kmps$})\hspace{-2mm} \\

\hline
A       &1.270 & 0.004  & 4.9  $\times$ 4.1  & -70 & 100 & 90      \\
A$_1$	&1.234 & -0.025 & 5.9  $\times$ 2.7  & -85 & 90  & 50--70  \\
A$_2$	&1.244 & 0.038  & 2.7  $\times$ 2.2  & -70 & 120 & 50      \\
B     	&1.255 & -0.086 & 9.1  $\times$ 4.5 &  70 & 80 & 70--100   \\
C       &1.260 & 0.080  & 4.6  $\times$ 3.8 & -45 & 90 & 60--100   \\
C$_1$	&1.232 & 0.079  & 4.6  $\times$ 2.9 & -65 & 120 & 50--90   \\
C$_2$	&1.238 & 0.113  & 3.0  $\times$ 3.0 &   0 & 180 & 50--100  \\
C$_3$	&1.263 & 0.115  & 3.3  $\times$ 2.0 & -45 & 100 & 50--70   \\
C$_4$	&1.291 & 0.051  & 3.3  $\times$ 2.9 & -80 & 120 & 60--80   \\
\hline
\end{tabular}

\end{center} \end{table}



\begin{figure*}[bthp]
\begin{center}
     \FigureFile(160mm,150mm){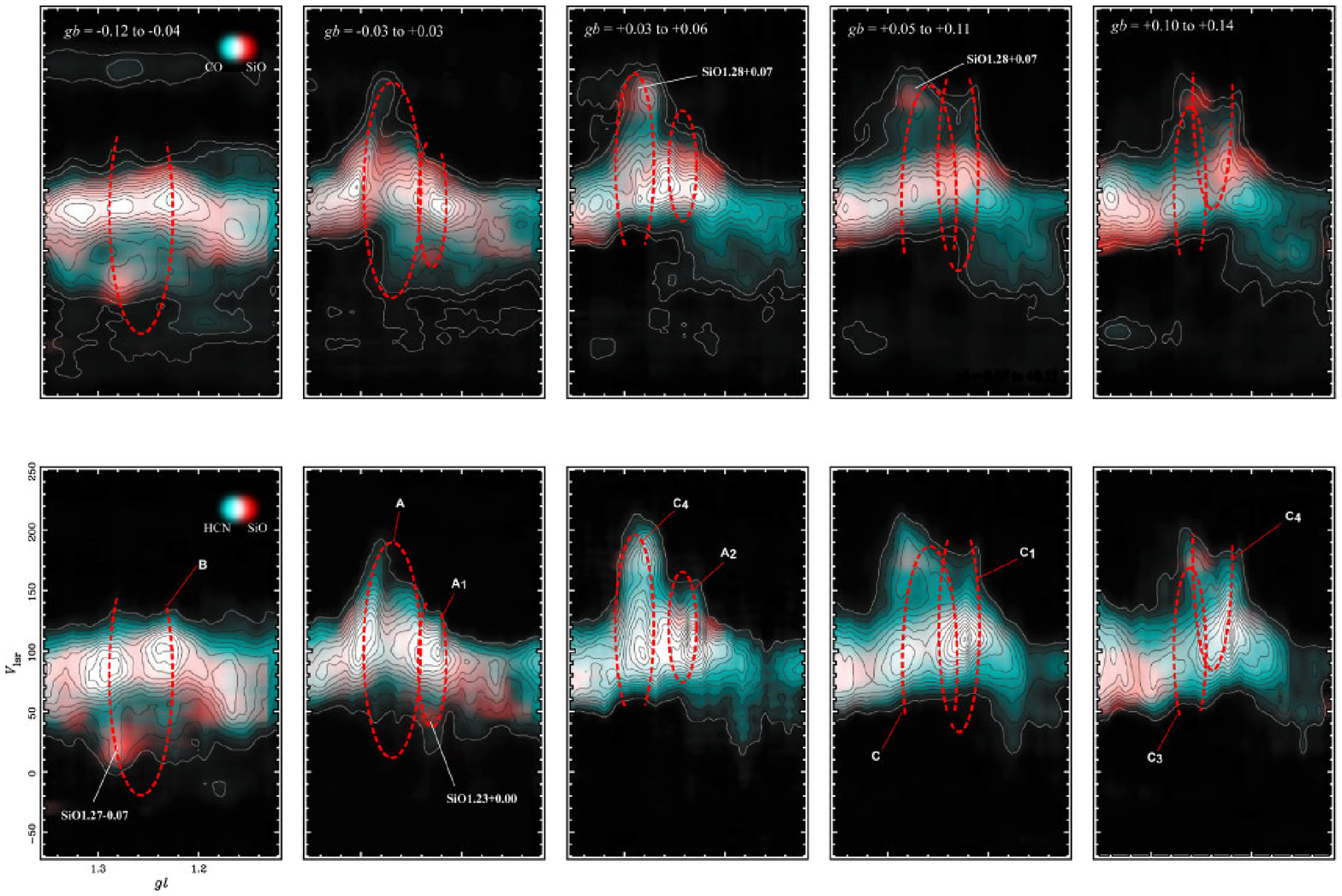}
\end{center}
\caption{
Longitude-velocity maps of CO $\JJ{1}{0}$ and HCN$\JJ{1}{0}$ line intensities (contours and gluegreen color maps).  
Contour levels are $4, 6, 8, 10 ..., \kelvin$ for CO and $0.5, 0.75, 1.0, 1.25 ..., \kelvin$ for HCN.  
SiO line intensity is represented by red color maps. }
     \label{Fig20}
\end{figure*}


\subsubsection {Shell A}
In Fig.\ref{Fig20} {\it l-V} maps of CO, HCN, $\HCOp{}$ and SiO $\JJ{1}{0}$ are presented.  
The shell A appears in the panel of $-0.03^\circ < b < +0.03^\circ$ in the figure.  
The shell A (the major shell) has a well-defined entity of expanding shell.  
The ``positive'' side of the expanding motion (i.e. $\vlsr > 100\ \kmps$) is clearly traced by a broad-velocity-width wing at $1.28^\circ < l <  1.29^\circ$ seen in the both CO and HCN maps.  
The HCN maps show that the systemic velocity of the shell is $\sim 100\ \kmps$ where the radius reaches a maximum.  
The HCN emission on the ``negative'' velocity side ($\vlsr < 100\ \kmps$) is less prominent, while a faint broad-velocity-width feature can be seen at $l \sim 1.25^\circ$ in the CO maps.  
This feature obeys the same kinematics of the Shell A determined in the positive-velocity emission, having a systemic velocity of $\sim 100\ \kmps$ and an expansion velocity of $90\ \kmps$, thereby it can be understood as a negative-velocity portion of the Shell A.  
Its faintness in the HCN line indicates low density.  

We also identified two minor shells (shell A$_1$ and A$_2$) adjacent to the shell A,  both of which are unrecognizable in the CO channel maps.  
The shell A$_1$ appears most clearly seen at $\vlsr = 120\ \kmps$ in the HCN map, having a shape elongated perpendicular to the Galactic plane.  
This shell accompanies several broad-velocity-width features in the Galactic west, $l\sim 1.23^\circ$, indicating that this shell is also expanding.  
The shell A$_2$ has a radius small ($\sim 40''$) compared to the other shells.  
This shell overlaps with the shell A on the plane of the sky in the velocity range from $80$ to $120\ \kmps$, but leaves the Shell A with increasing velocity.  
Only positive side of the expanding motion is traced by a broad-velocity-width feature at $(l, b) = (1.23^\circ, 0.05^\circ)$.  

We found compact SiO clumps well separated from the main velocity component ($\vlsr \sim 100\ \kmps$) on the {\it l-V} maps.  
These isolated SiO clumps should be related to the expanding shells, being located near the high/low velocity ends of broad-velocity-width features in many cases.  
The isolated SiO clump, SiO1.23+0.00, is located near the low velocity end of the shell A$_1$.  
The SiO clump SiO1.28+0.07 is also near the northern edge of the shell A.   
Unlike other two shells in the region, no isolated SiO feature is found to be associated to the shell A$_2$.

\subsubsection {Shell B}
The shell B has a shape elongated along the Galactic plane, with an aspect ratio $\sim 2$.  
The shell structure is subtly seen in the CO maps at its high velocity end.  In the HCN maps an arch consisting the upper half of the shell is clearly seen in $(l, b) = (1.25^\circ, -0.05^\circ)$ and $\vlsr = 100-120\ \kmps$, although the lower half of the shell is rather unclear.   

We can trace the expanding motion of the shell in the $-0.12 < b < -0.04$ panel of the $l-b$ maps.  
A prominent broad-velocity-width feature from $\sim 0\ \kmps$ to $\sim150\ \kmps$ is seen at the left side edge ($l \sim 1.28^\circ$) of the shell, which is also spatially elongated along the edge of the shell from $b=-0.09^\circ$ to $-0.04^\circ$.  
The HCN emission in the opposite side ($l\sim1.25^\circ$) of the shell B has also large velocity width $\sim 100\ \kmps$, although the velocity extent is somewhat smaller.

We see a SiO feature in the Galactic eastern edge of the shell B near its low velocity end, isolated from the main velocity component of the $l=1.3^\circ$ complex.  
This SiO feature has a CO counterpart.

\subsubsection {Shell C}
\citet{oka01} have found one expanding shell (the 'minor shell') in the northern part of CO1.27+0.01.  
The HCN and $\HCOp{}$ maps has more complicated structure than the CO map so that the kinematics cannot be described by a simple expanding motion.  
Especially in the velocity channel maps at $\vlsr = 140$-$200\ \kmps$, we can find a complex mixture of a number of arcs and filaments.  
We have identified 6 expanding shells in the region.
Prominent features are the shells C and C$_1$ (the minor shell).  
Although the position of the shell C in the $l-b$ plane is close to C$_1$, but is offset by $\sim +2'$ toward the Galactic east.  
The shape of the shell C is also different from C$_1$, having an aspect ratio close to 1, whereas C$_1$ has more elongated shape.  

Isolated SiO clump are associated also with the expanding shells in this area.  
SiO1.28+0.07 is the largest isolated SiO clump in the whole complex, located at the high velocity end of the shells C and C$_4$.  
The clump is also adjacent to the northern edge of the shell A.  
This SiO clump has a HCN counterpart, which is also isolated in velocity from the main component of the $l=1.3^\circ$ complex.

\subsection{HCN/HCO$^+$ Ratio}
\begin{figure}[bt]

\begin{center}
\FigureFile(60mm,60mm){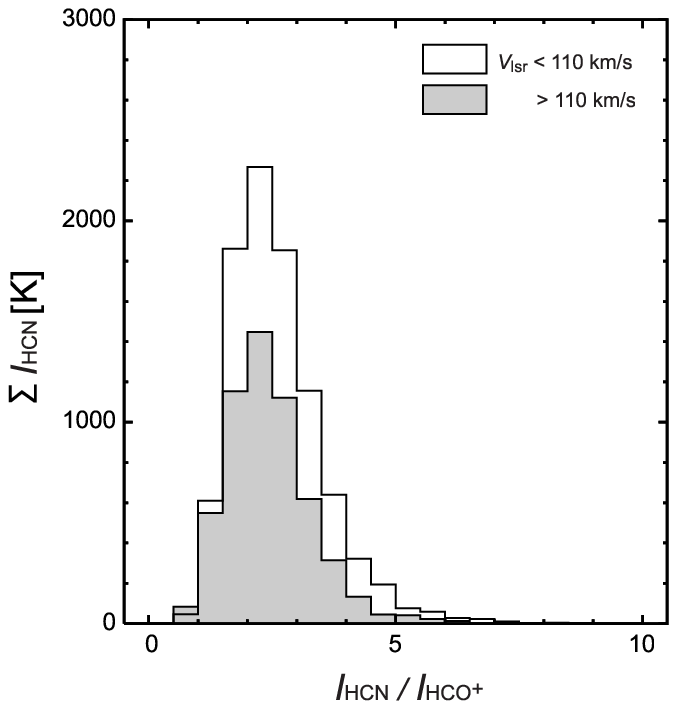}
\end{center}

\caption{Frequency histogram of HCN/$\HCOp{}$ intensity ratio. Frequency is weighted by HCN intensity.
}\label{Fig26}
\end{figure}

The HCN/$\HCOp{}$ ratios observed in the $\thecomplex$ complex are found to be higher by a factor of a few than those observed in typical molecular clouds in the Galactic disk ($\sim 1$; \cite{nym83, vw83, bla87}).
Figure \ref{Fig26} is a frequency histogram of the HCN/$\HCOp{}$ intensity ratio weighted by the HCN intensity.
The bulk of the gas has a ratio larger than unity.
Averaged intensity ratio is $2.3$, being quite similar to that measured toward the Circumnuclear Disk of our Galaxy\citep{chr05}. 
No systematic difference is seen between the ratios in the high velocity ($\vlsr > 110\ \kmps$) and low velocity ($\vlsr < 110\ \kmps$) components.

\section{DISCUSSION}
\subsection{Physical Conditions : Turbulent and Quiescent Gas}\label{SUBSECT_4_1}
It is found that the CO $\JJ{3}{2}$, HCN, $\HCOp{}$ and SiO intensities are enhanced in the high velocity gas ($\vlsr > 110\ \kmps$).  
The expanding shells are more prominent in this high-velocity gas, but in the low velocity range ($\vlsr < 110\ \kmps$) the spatial distribution of the CO emission becomes featureless.  
These results give the impression that the there are different types of gas with different kinematics and physical conditions.

\begin{figure}[bt]

\begin{center}
\FigureFile(80mm,80mm){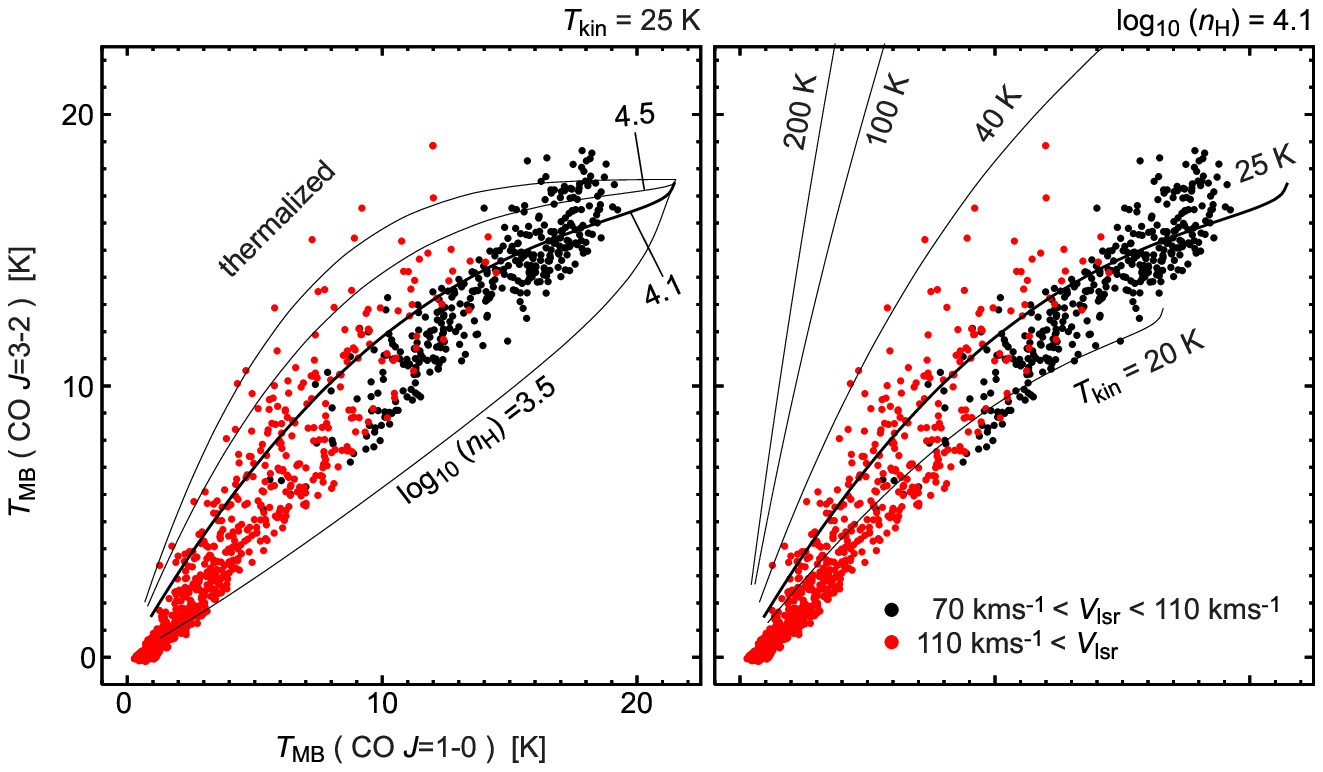}
\\ 
\FigureFile(80mm,80mm){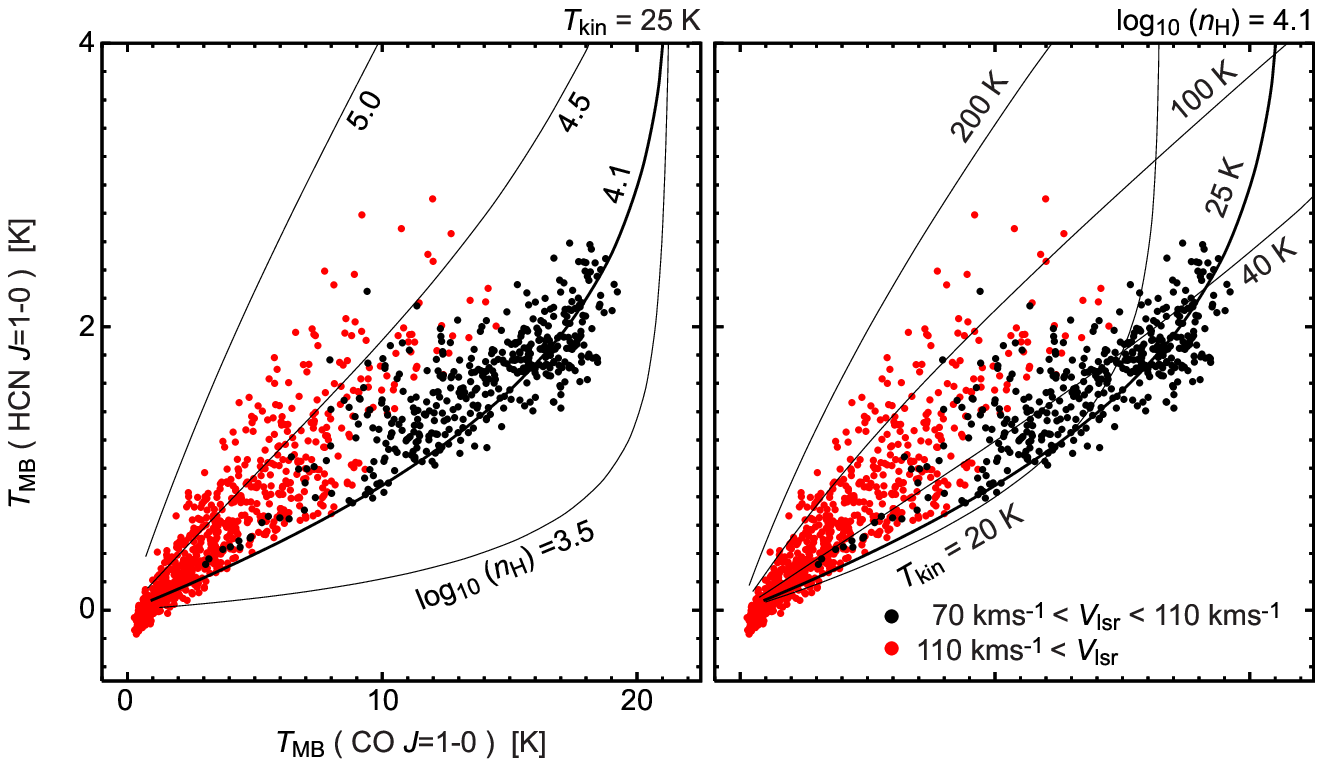}
\end{center}

\caption{
Calculated CO $\JJ{3}{2}$ (top) and HCN $\JJ{1}{0}$ (bottom) intensities plotted against 
CO $\JJ{1}{0}$ intensity for $\Tkin = 25\ \kelvin$ and various $\nH$ (left) and for $\nH = 10^{4.1}\ \pcc$ and various $\Tkin$ (right).  
Observed intensities are plotted separately for the high velocity ($\vlsr > 110\ \kmps$) and low velocity ($\vlsr < 110\ \kmps$).   
Relative HCN abundance to CO are assumed to be $10^{-3}$.
}\label{Fig3}
%




\begin{center}
\FigureFile(80mm,80mm){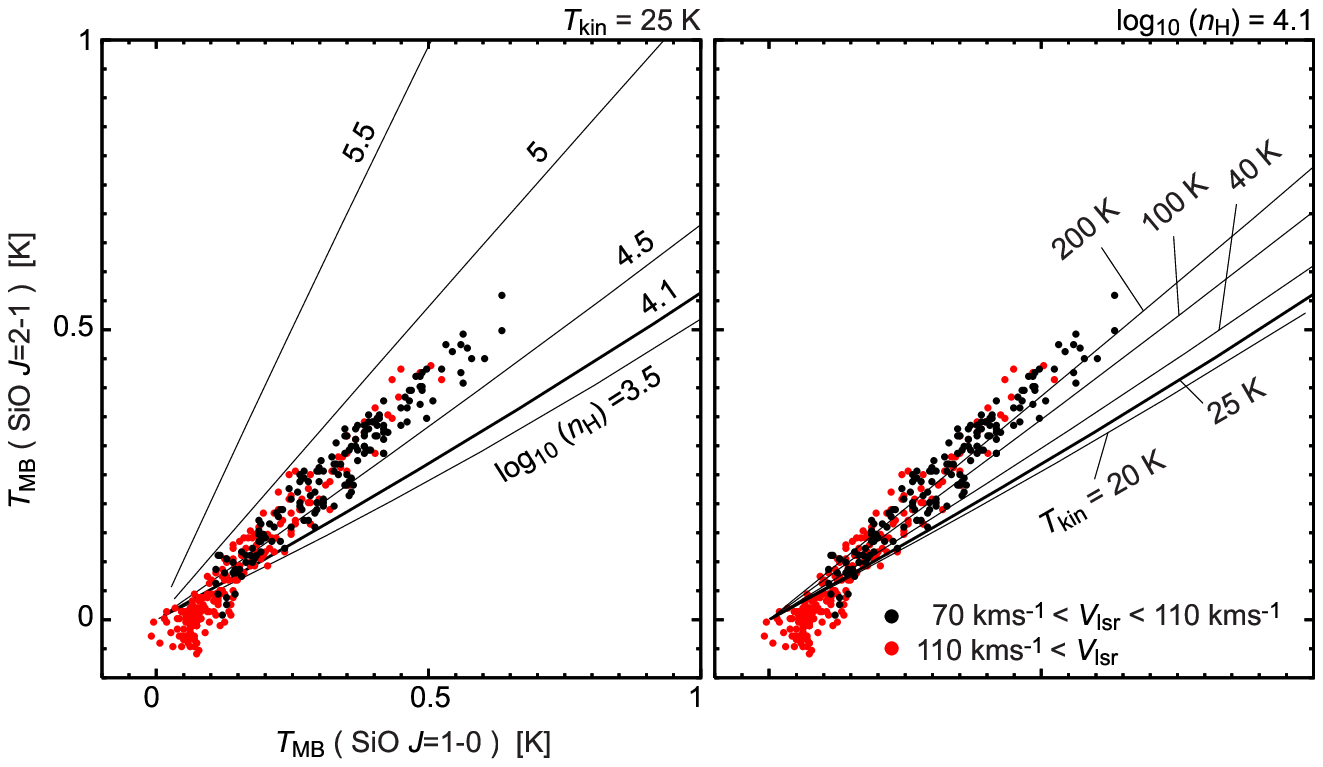}
\end{center}

\caption{Calculated SiO $\JJ{2}{1}$ intensity plotted against 
SiO $\JJ{1}{0}$ intensity for $\Tkin = 25\ \kelvin$ and various $\nH$ (left) and for $\nH = 10^{4.1}\ \pcc$ and various $\Tkin$ (right).  
Observed intensities are plotted separately for the high velocity ($\vlsr > 110\ \kmps$) and low velocity ($\vlsr < 110\ \kmps$). 
}\label{Fig9}

\end{figure}

We evaluated the dependence of the CO $\JJ{3}{2}$ and HCN $\JJ{1}{0}$ intensities on kinetic temperature and density by employing an LVG approximation.  
The parameters were kinetic temperature ($\Tkin$), hydrogen density ($\nH$) and CO column density per unit velocity width ($\dColumn{\CO{}{}}$).  
We assumed that $\XHCN = 10^{-3}$.
Intensities of HCN and CO $\JJ{3}{2}$ were calculated for $\Tkin = 20$-$200,\kelvin$ and $\nH = 10^{3.5}$-$10^5\ \pcc$ with varying $\dColumn{\CO{}{}}$.  
Figure \ref{Fig3} shows the calculated intensities plotted against CO $\JJ{1}{0}$ intensity.  
The observed intensities are also plotted for the high velocity ($\vlsr > 110\ \kmps$) and for low velocity ($\vlsr = 70\mbox{-}110\ \kmps$) separately.  

The results of calculations show that both the CO $\JJ{3}{2}$/$\JJ{1}{0}$ and HCN/CO intensity ratios become higher at higher density and at higher temperature for a wide parameter range.  
An exception to this trend is the case that the CO $\JJ{1}{0}$ becomes optically thick at $\Tkin = 20,\ 25\ \kelvin$ and the HCN $\JJ{1}{0}$/CO $\JJ{1}{0}$ ratio becomes larger at lower temperature.  
But this case is unlikely because the observed CO $\JJ{3}{2}$ / $\JJ{1}{0}$ ratios are higher than 1 in many data points, indicating that the CO emission in the high velocity gas are generally optically thin.  
The observed CO and HCN intensities  for the low velocity gas are well fitted by parameters $\Tkin = 25\ \kelvin$ and $\nH = 10^{4.1}\ \pcc$.  
The high velocity gas has higher density and/or higher temperature than those.  
It should be noticed that the CO $\JJ{3}{2}$ intensity is more sensitive to kinetic temperature than the HCN $\JJ{1}{0}$ intensity is.  
If density is fixed at $10^{4.1}\ \pcc$, the observed HCN intensity of the high velocity gas requires kinetic temperature of $\sim 100\ \kelvin$, which is too high to reproduce the observed CO $\JJ{3}{2}$ intensities.  
On the other hand, kinetic temperature should not be necessarily enhanced if density is $\gtrsim 10^{4.5}\ \pcc$ in the high velocity gas. This means that at least density should be enhanced in the high velocity gas.  

These results mean that moderate density gas ($\nH \sim 10^{4.1}\ \pcc$) and higher density gas ($\nH \gtrsim 10^{4.5}\ \pcc$) spread over the observed region.  
We also point out that that the observed SiO emission arises only from the latter component.  
Figure \ref{Fig9} shows observed SiO $\JJ{2}{1}$ intensity plotted against the $\SiO{}{}\,\JJ{1}{0}$ intensity separately for the high velocity and low velocity gas.  
Unlike the CO $\JJ{3}{2}$/CO $\JJ{1}{0}$ and HCN/CO $\JJ{0}{1}$ ratios, the SiO$\,\JJ{2}{1}$/SiO$\,\JJ{1}{0}$ ratio shows no systematic difference between the two velocity ranges.
Curves of calculated intensities are plotted in the figures.  
The SiO $\JJ{2}{1}$ / $\JJ{1}{0}$ ratio increases with density but only weakly depends on the kinetic temperature. 
The density should be $\gtrsim 10^{4.5}\ \pcc$ when the kinetic temperature is assumed to be $25\ \kelvin$, being consistent with the physical conditions of the high density gas.

The flat distribution of the SiO $\JJ{2}{1}$/$\JJ{1}{0}$ ratio means that the high density gas is also distributed in the low velocity range.  
In the low velocity range the high density and low density gas should be in the same line of sight.  
Our estimate of the physical parameters of the low density gas, $\Tkin = 25\ \kelvin$ and $\nH = 10^{4.1}\ \pcc$, should actually be considered to be average values of the both components.  
Therefore the density of the low density component should be lower than $10^{4.1}\ \pcc$.

These analyses show that the high density gas has nature of disturbed, turbulent gas, such as enhanced density and/or enhanced temperature and high SiO/CO intensity ratio, as well as large velocity dispersion.  
We can refer this component as the ``turbulent'' gas.  
The moderate density gas can be then considered as a ``quiescent'' gas, having a small velocity dispersion but lacks in SiO emission.  
The turbulent gas dominates the high velocity ($\vlsr > 110\ \kmps$) HCN and HCO$^+$ emission, while the quiescent gas confined in the low velocity ($\vlsr < 110\ \kmps$).  
This could be evidence that the shells in high velocities were formed via violent interaction of local explosion sources with surrounding ``quiescent'' material.  
The situation does not prefer the explanation that large-scale shock by cloud-cloud collision is responsible for the pervasive high SiO abundance and turbulent kinematics.


\subsection{SiO Abundance and Isolated SiO Features}\label{SUBSECT_4_2}
Figure \ref{Fig10} is a frequency histogram of SiO $\JJ{1}{0}$/$\CO{13}{}$ $\JJ{1}{0}$ intensity ratio weighted by the $\CO{13}{}$ intensity.  
Averaged ratio is 0.1 for $\vlsr < 110\ \kmps$ and 0.3 for $\vlsr > 110\ \kmps$.  
The [SiO]/[H] abundance ratio in the high velocity ($\vlsr > 100\ \kmps$) gas is estimated to be $10^{-9.3}$ when $\Tkin = 25\ \kelvin$ and $\nH = 10^{4.5}\ \pcc$, with assumption of $\CO{13}{}$/$\CO{12}{} = 24$\citep{lp90} and $n_{\rm H_2}/n_{\CO{13}{}} = 10^6$ \citep{lg89,lg90}.  
This is similar to the typical SiO abundance derived by \citet{hue98}.  
We should note that it is uncertain whether the SiO abundance is actually enhanced in the turbulent gas compared to that in the quiescent gas, because it is also possible that the density of the quiescent gas is not sufficiently high to excite the {\it J}=1 level of SiO.  If the density of the quiescent component is $10^{3.5}\ \pcc$, the SiO $\JJ{1}{0}$ / $\CO{13}{}\,\JJ{1}{0}$ intensity ratio can be smaller by an order of magnitude than that with density of $10^{4.5}\ \pcc$.  

We have found SiO emission features which are isolated in velocity from the main component at $\sim 100\ \kmps$ and spatially associated with the expanding shells.  
We listed 3 isolated SiO features in Table\ \ref{Tab3}. 
Spectra of SiO $\JJ{1}{0}$ and $\JJ{2}{1}$ lines toward the three isolated SiO features are displayed in Fig. \ref{Fig10} with $\CO{13}{}\,\JJ{1}{0}$ line.  
SiO1.28+0.07 and SiO1.28-0.07 have corresponding $\CO{13}{}\,\JJ{1}{0}$ peak at the same velocity as the SiO feature.  On the other hand the SiO1.23+0.00 does not have distinct $\CO{13}{}$ emission peak at the corresponding velocity, indicating that this is a shocked clump which overlaps in velocity with the quiescent gas. 

The SiO $\JJ{1}{0}$/$\CO{13}{}\,\JJ{1}{0}$ intensity ratios of the isolated SiO features are generally higher than those of the $100\ \kmps$ component.  
Especially the ratio at SiO1.28+0.07, 0.60, is higher than the typical value for the turbulent component by a factor of $\sim 2$.  
The SiO $\JJ{2}{1}$/$\JJ{1}{0}$ ratio is similar to the typical ratio in the region, $\sim 1$, indicating that the density and kinetic temperature are not largely different from those of the other part of the turbulent gas.  
Therefore the intense emission at SiO1.28+0.07 is considered as due to high SiO abundance.  
Estimated [SiO]/[CO] and [SiO]/[H] abundance ratios are $10^{-4.2}$ and $10^{-9.1}$, respectively.  
One possible explanation for the high SiO abundance of SiO1.28+0.07 is that this feature was formed relatively recently.  
SiO molecule is considered to be re-depleted onto the grain surface in the post-shock gas, and therefore the high SiO abundance in gas-phase could mean a shock has passed in recent past.  


\begin{table*} [bthp]
\caption{ISOLATED SiO CLOUDS}\label{Tab3}
\begin{center}
\small
\begin{tabular}{lcccccccccc}
\hline\hline
     & 
$l$ & 
$b$ & 
\multicolumn{2}{c}{\it isolated feature} & &
\multicolumn{2}{c}{\it 100 $km\,s^{-1}$ component} \\ \cline {4-5}\cline {7-8}
&&&
$\vlsr$ & \begin{minipage}{15mm}\vspace{1mm}$\frac{{\rm SiO}\ J=1-0}{{^{13}\rm CO}\ J=1-0}$\end{minipage} &&
$\vlsr$ & \begin{minipage}{15mm}\vspace{1mm}$\frac{{\rm SiO}\ J=1-0}{{^{13}\rm CO}\ J=1-0}$\end{minipage} \\
&
{\footnotesize (deg)} & 
{\footnotesize (deg)} & 
{\footnotesize $\rm(km\,s^{-1})$} &&&
{\footnotesize $\rm(km\,s^{-1})$} && 
\\
\hline
SiO1.28--0.07........... & 1.28 & -0.07 & 30 & 0.20 && 100 & 0.15\\
SiO1.23+0.00...........  & 1.23 & 0.00  & 60  & 0.21 && 90 & 0.14\\
SiO1.28+0.07...........  & 1.28 & 0.07  & 190 & 0.60 && 110 & 0.24\\
\hline
\end{tabular}

\end{center} \end{table*}


\subsection{Mass, Energy and Kinematical Age}

We derived the mass ($M$) and kinetic energy ($K$) of the shells.
The masses can be derived by summing up column density of the gas associated to each shell.
However, the entire complex is filled with shells, making it difficult to separate individual shell in the data cube.  
We adopt an approximation that all of the turbulent gas is associated with either of the 9 shells, and that the mass of expanding shell is proportional to the volume swept up by the shell.

Since the SiO emission traces the turbulent gas only, the ratio of the turbulent gas mass ($M_{\rm tur}$) to the total mass ($M_{\rm tot}$) is roughly estimated by the SiO $\JJ{1}{0}$/$\CO{13}{}$ $\JJ{1}{0}$ intensity ratio as follows: 
\begin{eqnarray}
\frac{M_{\rm tur}}{M_{\rm tot}} &=& \frac{\sum \Tmb\left(\rm SiO\right)}{0.3\ \sum \Tmb\left(\CO{13}{}\right)}\sim10^{-0.4}\ , 
\end{eqnarray}
where $\Tmb\left(\CO{13}{},\ {\rm SiO}\right)$ are the $\CO{13}{}$ $\JJ{1}{0}$ and SiO $\JJ{1}{0}$ intensities, respectively, and the summations are taken over velocities $20$-$200\ \kmps$ and the entire spatial coverage of SiO data.  
The total mass is estimated to be $M_{\rm tot}=10^{6.3}\ \Msol$ from the CO column density summed up over the region $1.19^\circ < l < 1.33^\circ$, $-0.20 < b < 0.17^\circ$.  The CO column density was calculated from the CO $\JJ{1}{0}$ and $\CO{13}{}\,\JJ{1}{0}$ data with the LTE assumption.  
We have adopted $\CO{13}{}$/$\rm H_2 = 10^{-6}$ and $\CO{}{}$/$\CO{13}{} = 24$.  
The turbulent gas mass is estimated to be $M_{\rm tur} = 10^{5.9}\ \Msol$.  

The volume swept by each shell, $V$, is assumed to be proportional to $\left(R_x\times R_y\right)^{1.5}$.  
Then the kinetic energy of the shell ($K$) is calculated by 
\begin{eqnarray}
K &=& \frac{1}{2}\  M_{\rm tur} \frac{V}{\sum_{\rm i}\ V_{\rm i}} \times {v_{\rm exp}}^2 \ ,
\end{eqnarray}
where the summation is taken over all of the 9 expanding shells.  

We also derived the kinematical ages of the expanding shells defined as
\begin{eqnarray}
\tau &\equiv& \frac{R}{v_{\rm exp}} 
     \sim \frac{\sqrt{R_{\rm x}\cdot R_{\rm y}}}{v_{\rm exp}}\ .
\end{eqnarray}
The value of $\tau$ becomes equal to the age if the shell has been expanding in a constant velocity.

We listed $M$, $K$ and $\tau$ in Table.\ref{Tab4}.  
The typical mass and energy are $10^{4.3-5.5}\ \Msol$ and $10^{50.7-52.5}\ \erg$, respectively.  
The total kinetic energy of the turbulent gas is $\sim 10^{52.7\mbox{--}53.0}\ \erg$.  
Note that the total mass and kinetic energy of the turbulent gas is not largely dependent on the identification of the shell, as far as the expansion velocity is $\sim 50\mbox{--}100\ \kmps$.  
The typical kinematical age is $10^{4.5\mbox{--}5.3}$ yr.  
This timescale is consistent with that SiO is still abundant in the turbulent gas, since the depletion time scale of SiO molecule onto grain is sufficiently long, $\sim 10^6$ yr \citep{rw96}. 

Uncertainties in our estimates of mass and kinetic energy mainly come from the uncertainty in the $\CO{13}{}$/$\rm H_2$  and CO/$\CO{13}{}$ abundance ratios.  
Validity of our procedure could be checked as follows.  
By assuming that the volume $\sum V_{\rm i}$ was originally filled with uniform density gas before swept up by the shells to the current morphology, the original density is derived to be $1.1\times10^4\ \pcc$ from the mass and volume of the shells.  
This is consistent with the typical density in the CMZ ($\sim 10^4\ \pcc$).  
Our HCN data also indicate that the quiescent gas has a density of $<10^{4.1}\ \pcc$ (Sections \ref{SUBSECT_4_1} and \ref{SUBSECT_4_2}).  
We can say that our mass estimate is consistent within a factor of a few with the densities derived with independent procedures.


\begin{figure}[btp]
\begin{center}
\FigureFile(60mm,60mm){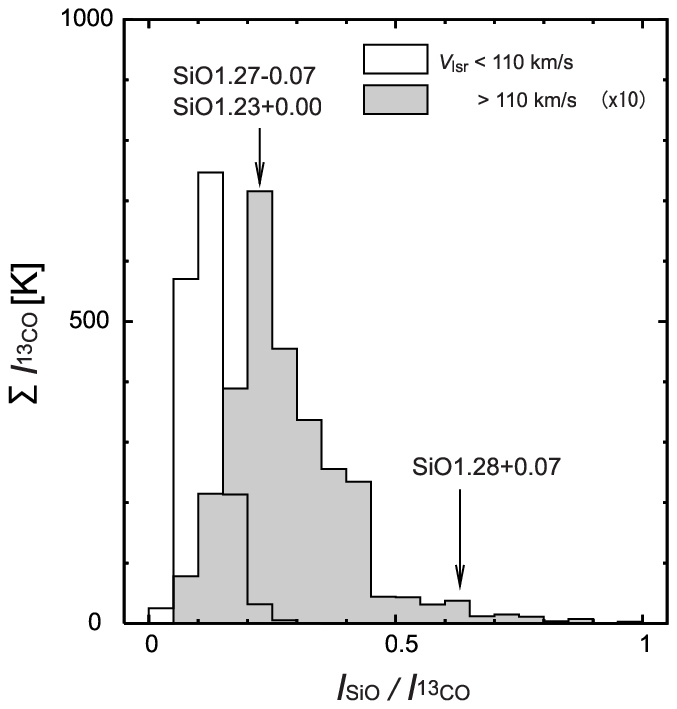}
\caption{Frequency histogram of SiO $\JJ{1}{0}$/$\CO{13}{}\,\JJ{1}{0}$ intensity ratio. The frequency is weighted by the $\CO{13}{}$ intensity. 
} \label{Fig10}
\end{center}
\end{figure}

\begin{figure*}[btp]
\begin{center}
\FigureFile(100mm, 100mm){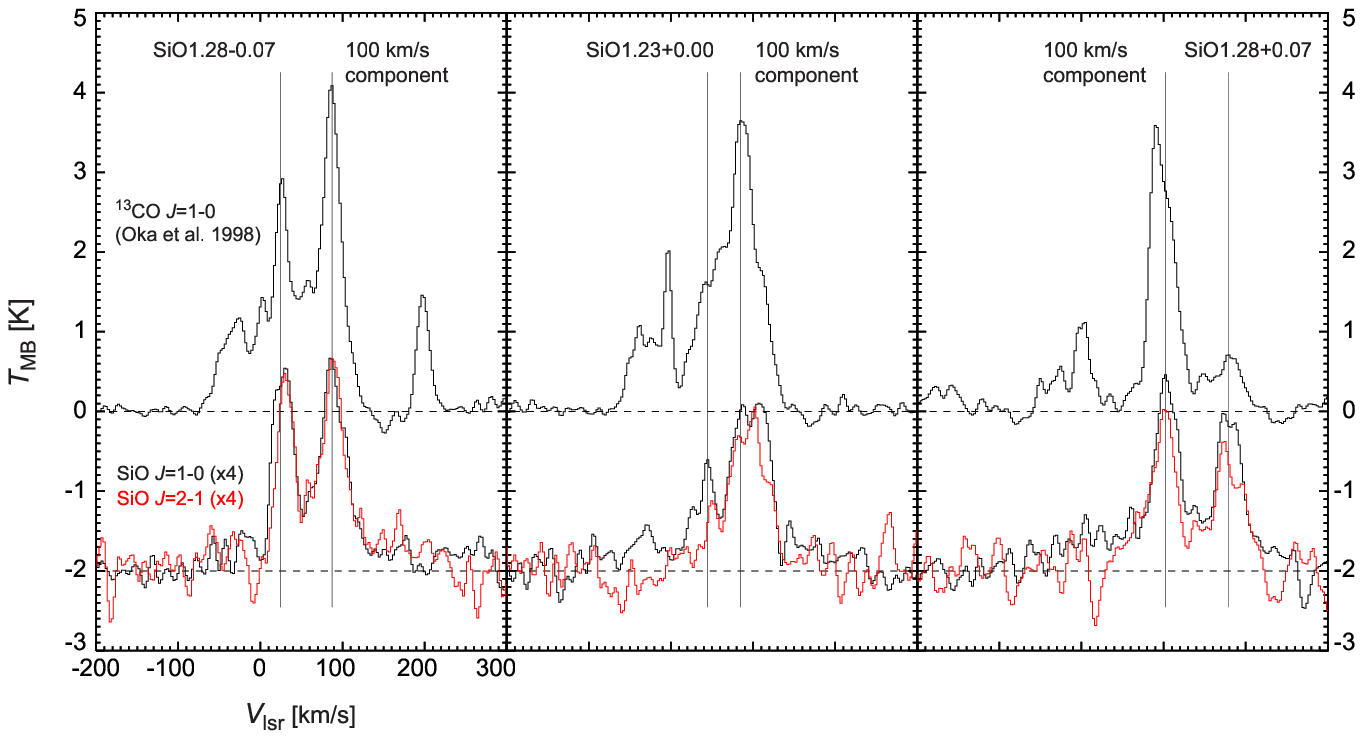}
\caption{Spectra of $\CO{13}{}\,\JJ{1}{0}$ \citep{oka98}, SiO $\JJ{1}{0}$ and $\JJ{2}{1}$ lines toward the positions of isolated SiO features.
} \label{Fig10}

\end{center}
\end{figure*} 


\subsection{The Origin of the Turbulent Gas : Proto-Superbubble?}

What is the source of huge kinetic energy of expanding shells?.  
As discussed in \citet{oka01} the stellar wind from Wolf-Rayet stars is not sufficient to furnish the kinetic energy even for a single shell.  
Promising candidates may be supernova (SN) or hypernova (HN) explosions.  
The total kinetic energy of the turbulent gas, $\sim 10^{52.7\mbox{--}53.0}\ \erg$, corresponds to 50--100 SN explosions.  
Each of the three major shells A, B and C has kinetic energy of $10^{52}\ \erg$, which requires several SN explosions or a single HN explosion.  
Other minor shells have smaller energy, $\sim 10^{51}\ \erg$, which can be furnished by a single SN.  
Therefore the required number of SN and HN explosions can be estimated to be $\sim 10$--$100$.  
Combined with the kinematical age of the shells, $\sim 10^5$ yr, the SN rate should be $10^{-3\mbox{--}-4}\ {\rm yr^{-1}}$.  

A massive stellar cluster could be the progenitor of the successive SN/HN explosions.  
Duration of SN explosions in a cluster with a single population stars is $\sim 10^{7.6}\ \yr$, roughly the lifetime of a star of $8 \Msol$ \citep{lei99}.  
If the cluster was formed with timescale sufficiently shorter than that, the SN rate is determined by the cluster mass.  
The required cluster mass is then estimated to be $10^{5\mbox{--}6}\ \Msol$ by assuming the Salpeter IMF with the lower/higher mass cutoffs of $1\ \Msol$/$100\ \Msol$.  
Another extreme case is that the star formation continued for longer than $10^{7.6}\ \yr$.  
In this case SN rate is determined by the star formation rate (SFR).  
Then the SFR is estimated to be $\sim 10^{-1\mbox{--}-2} \ \Msol \rm yr^{-1}$ with the Salpeter IMF.  
The total molecular mass in the observed area is $\sim 10^{6.3}\ \Msol$, and then the star formation efficiency (SFE) is $3\times 10^{-8\mbox{--}-9}$ yr$^{-1}$.  
This is similar to or larger than the averaged SFE of the entire CMZ, $5\times10^{-9}\ {\rm yr^{-1}}$, estimated from the 6 cm continuum flux \citep{gus89}.

It should be noticed that no sign of current massive star formation activity is found in the $\thecomplex$ complex.  
Although a large HII region, Sgr D, is located adjacent to the observed region, it is suspected of being in the Galactic disk \citep{meh98}.  
The lack of HII regions prefers that a massive cluster older than the typical lifetime of O-type stars ($\sim 10^6$ yr) is the progenitor of the successive SN/HN explosions.  
The model of \citet{lei99} shows that the SN rate remains constant for $\sim 10^{7.6}$ yr after O-type stars vanished at $t = 10^{6.8}\ \yr$. 
We did not find any expanding shell with the kinematical age older than $10^{5.3}$ yr.  
This should be due to a selection effect: shells produced by SNe older than $\sim 10^6$ yr would be too large or too fragmented to be recognizable.  

If the gathering of expanding shells in the $\thecomplex$ complex is due to multiple SN/HN explosions, it would evolve into molecular superbubbles seen in starburst galaxies (eg. \cite{sak06}).  
The situation could be described that {\it the $l=1.3^\circ$ complex contains a molecular superbubble in the early formation stage}.  
This proto-superbubble may be originated by a massive cluster formation which took place $10^{6.8\mbox{--}7.6}\ \yr$ ago.  
This could be evidence that the complex was a site of a burst-like star formation which is similar to that presently ongoing in the Sgr B2 \citep{mpg95, gau95, dep96}.   

However, we should note that no nonthermal radio continuum source indicating presence of SNR is associated to the molecular shells \citep{lis92, yz04}.  
Diffuse nonthermal emission extends over the entire CMZ \citep{lar05}, but the authors argue that much lower SN rate of $10^{-5}\rm\ yr^{-1}$ is sufficient to produce the observed intensity under the weak magnetic field ($\sim 10\ \mu{\rm G}$).  
This implies that, in the CMZ, particle acceleration in a SN blast wave is inefficient.  The mechanism which reduces acceleration efficiency is still an open question.   


\begin{table}[bt]
\caption{MASS, KINETIC ENERGY AND KINEMATICAL AGE}\label{Tab4}
\begin{center}
\pagestyle{empty}
\small
\begin{tabular}{lccc}
\hline\hline
 & 
$\log_{10}M$ &
$\log_{10}K$ &
$\log_{10}\tau$ \\ 
 &  
{\footnotesize($\Msol$}) &
{\footnotesize($\erg$})  &
{\footnotesize(yr}) \\

\hline
A     & 5.1 & 52.0 & 4.6\\
A$_1$ & 4.9 & 51.3--51.6 & 4.8--5.1\\
A$_2$ & 4.3 & 50.7 & 4.7\\
B     & 5.5 & 52.2--52.5 & 4.8--5.1\\
C     & 5.0 & 51.8--51.9 & 4.6--4.7\\
C$_1$ & 4.8 & 51.2--51.7 & 4.6--5.1\\
C$_2$ & 4.5 & 50.9--51.5 & 4.5--5.3\\
C$_3$ & 4.3 & 50.7--51.0 & 4.6--4.9\\
C$_4$ & 4.6 & 51.2--51.4 & 4.6--4.8\\
\hline
$turbulent$ \\
$\hspace{10mm}gas$\hspace{-5mm} & 5.9 & 52.6--53.0 & \\
$total\ mass$                    & 6.3 & & \\
\hline
\end{tabular}

\end{center} \end{table}


\section{SUMMARY}
We made high resolution observations of CO $\JJ{1}{0}$, HCN $\JJ{1}{0}$, $\HCOp{}\,\JJ{1}{0}$, SiO $\JJ{1}{0}$ and $\JJ{2}{1}$ toward the $\thecomplex$ complex in the Galactic Center, as a follow-up of the CO $\JJ{3}{2}$ survey carried out with the ASTE 10\ m telescope (Paper I).  The main results are summarized as follows:  
\begin{enumerate}
\item 
The complex is found to be rich in expanding shells and arcs of dense molecular gas.  We have identified 7 expanding shells in addition to the previously known two energetic shells.  
\item
The HCN/CO and $\HCOp{}{}$/CO ratios are enhanced by a factor of a few in the velocity range higher than $110\ \kmps$.  The ratios of SiO $\JJ{1}{0}$/$\CO{13}{}\,\JJ{1}{0}$ and CO $\JJ{3}{2}$/$\JJ{1}{0}$ are also enhanced in the high velocity.  
\item
LVG analyses indicate that there are two types of gas with different physical conditions and kinematics. The high-velocity `turbulent' gas has high density ($\nH\sim10^{4.5}\ \pcc$), high SiO/$\CO{13}{}$ intensity ratio and large velocity dispersion.  The bulk of gas is `quiescent', having  moderate density ($\nH\lesssim10^{4.1}\ \pcc$) and velocity width ($\sim 40\ \kmps$).  
\item
Compact SiO emission features isolated from the main velocity component ($\vlsr\sim 100\ \kmps$) are found near the high/low velocity ends of the expanding shells.  
The most prominent isolated SiO feature, SiO1.27+0.07, has an enhanced SiO abundance of $\sim 10^{-9.1}$.  It may be the site where a shock wave has passed relatively recently.    
\item
The typical HCN/$\HCOp{}$ intensity ratio is found to be 2.3, being higher than that in the Galactic disk region, $\sim 1$.
The ratio is close to the value measured toward the CND of our Galaxy.
\item
The kinetic energies of the expanding shells are of the order of $10^{51\mbox{--}52}$ erg.  The total kinetic energy of the 9 expanding shells is $10^{52.6\mbox{--}53.0}$ erg, and the kinematical age is $\sim 10^5$ yr. 
Multiple SN/HN explosions at the rate of $10^{-3\mbox{--}-4}\rm yr^{-1}$ can furnish the large kinetic energy of the expanding shells.  
\item
The expanding shells as a whole may be in the early stage of superbubble formation.  
This proto-superbubble may be originated by a massive ($\sim 10^{5\mbox{--}6}\  \Msol$) cluster formation or continuous star formation at the rate of $\sim 10^{-1\mbox{--}-2} \ \Msol \rm yr^{-1}$.
The absence of developed HII regions in the complex could indicate those activities took place $10^{6.8\mbox{--}7.6}\ \yr$ ago.
\end{enumerate}

\end{document}